\documentclass[11pt]{article}
\pdfoutput=1

\usepackage[margin=1in, paperwidth=8.5in, paperheight=11in]{geometry}

\usepackage{graphicx}
\usepackage{float}
\usepackage{caption}
\usepackage{subcaption}
\usepackage{fp}
\usepackage{pgfplots}
\pgfplotsset{compat=1.13}

\usepackage{tikz}

\usepackage{verbatim}

\usepackage{booktabs}

\usepackage[numbers]{natbib}
\usepackage{url}

\begin{document}

\title{ A Method of Finding a Lower Energy Solution to a QUBO/Ising Objective Function }
\author{John E. Dorband\\
	Department of Computer Science and Electrical Engineering\\
	University of Maryland, Baltimore County\\
	Maryland, USA\\
	\texttt{dorband@umbc.edu}}
\date{\today}
\maketitle

\begin{abstract}

A new method to find a lower energy solution to a QUBO/Ising objective function will be presented in this paper.
It is applied to samples returned from the D-Wave for various example cases.
This method, multi-qubit correction (MQC), creates a sample with an equal-to or less-than energy than any
of the D-wave samples used to create it. 
The method will be detailed and the results of 3 uses cases will be given to demonstrate its merit.

\end{abstract}

\section{Introduction}\label{sec:intro}

The D-Wave\citep{Dwave13} is an adiabatic quantum computer\citep{Farhi00,Giuseppe08}.
The problem class that is addressed by the D-Wave is based on the Ising model objective function, $F$:
\begin{equation}\label{eq:obfunc}
F = {\sum\limits_i a_i q_i + \sum\limits_i \sum\limits_j b_{ij} q_i q_j}
\end{equation}
where $q_i\in\{-1,1\}$ are the qubit values returned by the D-Wave, and $a_i\in[-2,2]$ and $b_{ij}\in[-1,1]$ are 
the coefficients given to the D-Wave associated with the qubits and the qubit couplers respectively. 

There has been extensive work done on quantum error correction, in particular on error correction of 
quantum annealing computations\citep{Young2013,Pudenz2014,Vinci2017}. 
The D-Wave is also considered a quantum annealing computer. 
These works on quantum error correction focus on improving quantum architecture, 
finding better physical parameter adjustment,
and encoding additional physical qubits for error correction into the quantum problem to be solved.

The purpose of a quantum annealer is to find the absolute (global) minimum of the expression $F$,
which is NP-hard.
A simpler problem can be expressed by $F$, an NP-complete problem,
where the pattern of qubit values is what is important, not the global minimum.
An NP-complete problem solution can be quickly verified,
and yet finding the pattern of qubit values is intractable.
In either case, determining the correctness of the solution as it is being computed is extremely difficult.
The complexity of defining error correction of a system in which large numbers 
of qubits are interacting with each of over a very short period of time only exacerbate the problem of 
defining what exactly error correction is.
If one defines error correction as finding the best possible minimization of $F$, then it is little more than
a heuristic that may come close to the minimum, but one will never know if it has been reached.
Or if you define error correction as correcting the bit pattern before you've completed the computation,
when the problem is intractable and your computation takes only a single step, one has a dilemma.
This work focuses on a heuristic approach to improving the minimum value of $F$, in the expectation
that the resulting qubit value pattern is correct.

A D-Wave 2x can have as many as 1152 qubit coefficients and 3360 coupler coefficients.
A D-Wave 2000Q can have as many as 2048 qubit coefficients and 6016 coupler coefficients.
Therefore to utilize the D-Wave, one must come up with a set of coefficients and send them to the D-Wave.
The D-Wave then returns at least one set of qubit values. 
This set is referred to here as a sample.
A request may be for many more than one sample. 
A typical request here results in 1000 samples.
The D-Wave's purpose is to return the set of qubit values which minimize $F$. 
The value of the objective function, $F$, for a sample will be referred to as the sample's energy.
There can only be one global minimum energy, though there may be multiple samples with that global minimum energy.  
This global minimum energy corresponds to the ground state of the D-Wave for the given set of coefficients.  
The D-Wave often returns a non-minimum energy state due to inherent quantum/thermal noise in the system 
or the closeness of a large number of slightly higher energy 'active' states near the ground state.  
This leads to the question: if the D-Wave does not always return qubit values corresponding to the ground 
state what are the properties of the D-wave that can be depended upon to perform useful computations. 
Previously, the author addressed this issue by characterizing the behavior of the D-Wave with various test 
cases\citep{Dorband16}.

This paper addresses quantum annealer correction more from a results improvement perspective than a
qubit value correction perspective.
The heuristic approach defined in this paper uses a post-processing method performed classically on the quantum
annealer's returned results.
Thus the method may be performed on any quantum annealer independent of the quantum architecture
or any additional architectural or algorithmic improvements.
This work focuses on extracting whatever information possible from the quantum annealer's results 
that can be utilize to improve the final results.
The method's primary goal is to find the global minimum value of $F$.
It's secondary goal, in some cases it's a more important goal, 
is to discover the qubit value pattern that represents the global minimum value of $F$.

In the D-Wave improvement algorithm, single qubit correction(SQC), described in \citep{Dorband17}, 
each qubit was processed one at a time.
The influence of the qubit on $F$ was calculated and if the influence was negative, its value would be left
alone. 
But if the influence was positive, the value of the qubit would be negated, 
thus making the influence negative and lowering the overall sample energy. 
This is done for every sample individually, forcing the energy of each sample to a value of $F$
that is a minimum value but seldom the global minimum.

In this paper, a more robust algorithm will be presented, multi-qubit correction (MQC).
Instead of processing one qubit at a time on samples individually, groups of qubits are treated as units, 
tunnels, and samples are compared with each other to find these tunnels.
The term `tunnel' here is analogous to the concept of quantum tunnels as presented in \citep{Denchev15}.
In \citep{Denchev15}, a quantum tunnel is demonstrated as the property of the D-Wave to simultaneously flip a group
of qubit values thus changing a non-global minimum solution to a global minimum solution.

So a tunnel is defined here as a group qubits which has an influence value, $I$.
When all its qubit values are negated (flipped), they influence the
value of $F$ by either increasing it by $I$, decreasing it by $I$, or not changing it at all, $I=0$.
A positive influence, $I$, implies that the tunnel qubits should be flipped, to lower the value of $F$,
a negative influence means the the tunnel should be left as is, and
a zero influence allows the tunnel to change the sample to a new sample which has the same value of $F$
resulting in additional samples with the same energy.
In the case of a sample with globally minimum energy, a zero influence tunnel will give multiple samples
for the ground state solution.

The assumptions here is that 1) the D-Wave seldom find either the global minimum or a local minimum, 
2) that it always finds something near the ground state, and better yet 3) it finds components of the global 
minimum or tunnels that lead to the global minimum.

In this paper, a method, MQC, will be presented to find tunnels which will lead to the construction of
a more minimal solution and possibly the globally minimum solution.
Note that in many if not most cases, it is not provable that a solution is a global minimum.
It can only be shown that one sample represents a lower energy than another.
This paper will also present three use cases that demonstrate the merit of MQC.

\section{Multi-qubit Correction Method (MQC)}\label{sec:Method}

The objective of multi-qubit correction is to construct a sample that has globally
minimal energy, though not provably the global minimum.
This is done by constructing a new sample for every pair of samples, 
reducing the number of samples by half. 
This is repeated until only one sample is left.
Each constructed new sample will have an energy that is equal-to or less-than either of the two samples
that it was created from. 

\subsection{New sample construction}\label{sec:NewConstruct}

The construction of a new sample starts with two samples, $A$ and $B$.
$A$ and $B$ are compared to determine if qubits $q^A_i$ and $q^B_i$ are equal or not.
Two sets of qubit indices are created $i \in S$ if $q^A_i = q^B_i$ and $i \in D$ if $q^A_i \neq q^B_i$.
$D$ is a tunnel which represents the transformation of sample $A$ to sample $B$.
This in itself is rather useless. 
However it can be used to find sub-tunnels that may be used to transform
$A$ or $B$ into a sample with lower energy.
The next step is to find those sub-tunnels.
A sub-tunnel, $T$, of $D$ consist of all $q_i$ and $q_j$ where $b_{ij} \neq 0$ and $i,j \in D$.
Therefore $T$ is the closure of a set qubits connected transitively to each other, but not connected to
other qubits in $D$.
This results in a set of tunnels $T^k$ that have no non-zero connections between them.
All $T^k$ are independent of each other (Eq. \ref{eq:tunnelInter}).
There are no zero value couplers ($b_{ij}$) between any qubit in one tunnel and a qubit in another.
And yet the union of all tunnels is equal to the tunnel $D$ (Eq. \ref{eq:tunnelUnion}). 

\begin{equation}\label{eq:tunnelInter}
 {T^s} \cap {T^t} = \emptyset, \forall s,t
\end{equation}
 
\begin{equation}\label{eq:tunnelUnion}
\bigcup\limits_{k=1}^n T^k = D
\end{equation}

where $n$ is the number of tunnels found in $D$.
The influence of $T^k$ is its contribution to $A$ or $B$.
Note that the energy contribution of $T^k$ ( its influence ) to $A$ is the negative of its contribution
to $B$. 
The influence of $T^k$ relative to sample $A$ is as follows:
\begin{equation}\label{eq:TunnelInfluenceExp}
I^k_A = {\sum\limits_{i \in T^k} a_i q^A_i + \sum\limits_{i \in T^k} \sum\limits_{j \in S} b_{ij} q^A_i q_j }
\end{equation}
Note that $q_j$ is the same for both samples $A$ and $B$.
Also note that $\sum\limits_{i \in T^k} \sum\limits_{j \in T^k} b_{ij} q^A_i q^A_j$ is not included in
the expression above. 
This because it has the same value no matter whether the values of the qubits of $T^k$ are flipped or not
and does not effect the relative influence of $I^k_A$ and $I^k_B$.

The lowest energy common sample can be obtained from either sample $A$ or $B$.  
If the influence of a tunnel $T^k$ has a positive value relative to $A$, its qubit values are flipped
giving a new sample with a lower energy.
If this is done for all $T^k$ such that $I^k_A > 0$, the result is a new sample $C$ with a minimal energy relative to $A$.
If this is done for $B$, the same minimal energy sample $C$ is created.

\subsection{Sample Aggregation}\label{sec:T0}

As stated before, all the samples obtained from the physical quantum machine (eq. D-Wave) will be aggregated
into a single sample. 
The energy of this sample will be less-then or equal-to the lowest energy sample obtained from the quantum machine.
If $n$ samples were originally obtained from the quantum device, 
the first aggregation step will produce  $n/2$ new samples.
In $O(log(n))$ steps the $n$ samples will be reduced to one sample.

\section{Multi-qubit Correction Examples}\label{sec:Examples}

D-Wave sample correction is simple and can be easily incorporated into any algorithm that
has direct access to samples returned by the D-Wave.
In the following sub-sections, the results from 3 algorithms are presented which utilize corrected 
and uncorrected D-Wave samples.
The first example is a study of minimizing $F$ for a set of random values, qubit and coupler coefficients.
The second example expands on D-Wave characteristics shown in \citep{Dorband16}.
It shows the behavior of the qubit chain (virtual qubits) using uncorrected D-Wave samples, SQC corrected samples,
MQC corrected samples, and theoretically correct virtual qubits.
The third example shows how well a Chimera Boltzmann machine\citep{Dorband15} trained on 
hand written digits from the MNIST data set performs using uncorrected, SQC corrected, and MQC corrected 
D-Wave samples.

\subsection{Example 1: Objective function with random coefficients}\label{sec:RandomCoeff}

In this example, 1000 cases were created consisting of random values for all qubit and coupler 
coefficients. 
For each case 10,000 samples were requested from the D-Wave 2X at Burnaby, BC.
The 10,000 samples formed into 10 sample sub-sets of increasing size.
Each sub-set containing all the samples of the preceding set.
Starting with 1000 samples in the first sub-set and adding 1000 samples to each succeeding sub-set with
the 10th and last sub-set containing all 10,000 samples.
MQC was applied to each sub-set of samples for each case.

\begin{table}[H]
\begin{tabular}{|l|c|c|c|c|c|c|c|c|c|c|c|}
\hline
N & 0K & 1K & 2K & 3K & 4K & 5K & 6K & 7K & 8K & 9K & 10K      \\ \hline
Cases & 4  & 789  & 82 & 43 & 21 & 13 & 18 & 11 & 11 & 4 & 4   \\ \hline
\end{tabular}
\centering
\caption{Samples that reach final minimum within N samples.}
\label{table:RandomCases}
\end{table}


Table \ref{table:RandomCases} shows how many cases reached their MQC minimum sample by the 
$N^{th}$ set of samples and showed no improvement for any larger set of samples.
The 4 cases indicated in column 0K showed no improvement over the uncorrected D-Wave samples.
Indicating that MQC did not improve over the minimum D-Wave sample. 
Out of the 1000 cases MQC improved 996 cases, taking anywhere from 1000 to 10000 D-Wave samples to do it.

One can see in \citep{Dorband17} that when SQC is applied to a case with random coefficients just shifts
the distribution of sample energies toward a lower value.

Of the 1000 cases used in this example, SQC produced lower minimum energies than the D-Wave for 976 cases and MQC
produced lower minimum energies than SQC for 941 cases.
Of the cases that SQC or MQC did not return lower energies their energies were the same to within $10^{-14}$.

\subsection{Example 2: Virtual qubit characteristics}\label{sec:VirtualQubit}

The concept of a virtual qubit is a group of physical qubits that act as one.
Therefore if one of the qubits has a true value all the qubits should have a true value,
and if one of the qubits has a false value all the qubits should have a false value.
Since the Ising model is being assumed here, true is equivalent to 1 and false is equivalent to -1.
Virtual qubits are represented here with 12-bit chains of physical qubits.
Seldom however are the values of the qubits of a virtual qubit all true or all false.

\begin{figure}[H]
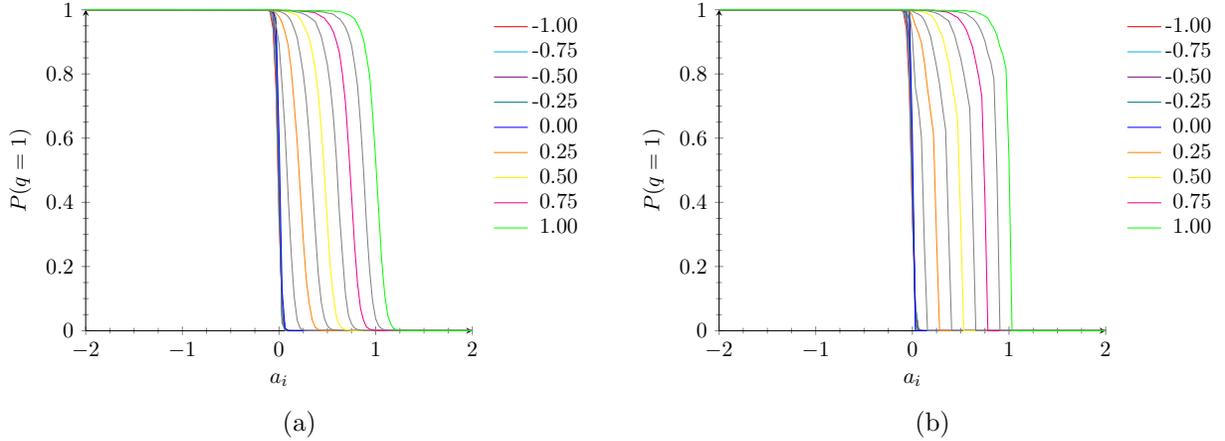

    \begin{subfigure}{0.49\textwidth}
	\scalebox{0.75}{
	    \begin{tikzpicture}
		\input{tab/chain_ISING_dwave_vote_129_017}
		  \begin{axis} [
      axis lines=left,
      xlabel={$a_i$},
      ylabel={$P(q=1)$},
      ymin=0, ymax=1,
      xmin=-2, xmax=2,
      minor x tick num=3,
      minor y tick num=3,
      legend cell align=right,
      legend style={ draw=none, at={(1.3,1.0)}, },
    ]
    \addplot[red]    table[ x index=1, y index=2,  ]                {\plotdata};
    \addplot[gray]   table[ x index=1, y index=3,  forget plot, ]   {\plotdata};
    \addplot[cyan]   table[ x index=1, y index=4,  ]                {\plotdata};
    \addplot[gray]   table[ x index=1, y index=5,  forget plot, ]   {\plotdata};
    \addplot[violet] table[ x index=1, y index=6,  ]                {\plotdata};
    \addplot[gray]   table[ x index=1, y index=7,  forget plot, ]   {\plotdata};
    \addplot[teal]   table[ x index=1, y index=8,  ]                {\plotdata};
    \addplot[gray]   table[ x index=1, y index=9,  forget plot, ]   {\plotdata};
    \addplot[blue]   table[ x index=1, y index=10, ]                {\plotdata};
    \addplot[gray]   table[ x index=1, y index=11, forget plot, ]   {\plotdata};
    \addplot[orange] table[ x index=1, y index=12, ]                {\plotdata};
    \addplot[gray]   table[ x index=1, y index=13, forget plot, ]   {\plotdata};
    \addplot[yellow] table[ x index=1, y index=14, ]                {\plotdata};
    \addplot[gray]   table[ x index=1, y index=15, forget plot, ]   {\plotdata};
    \addplot[magenta]table[ x index=1, y index=16, ]                {\plotdata};
    \addplot[gray]   table[ x index=1, y index=17, forget plot, ]   {\plotdata};
    \addplot[green]  table[ x index=1, y index=18, ]                {\plotdata};
    \legend{-1.00,-0.75,-0.50,-0.25,0.00,0.25,0.50,0.75,1.00}
  \end{axis}
	    \end{tikzpicture}
	}
	\caption{ } \label{fig:IsingRaw129}
    \end{subfigure}
    \hspace*{\fill} 
    \begin{subfigure}{0.49\textwidth}
	\scalebox{0.75}{
	    \begin{tikzpicture}
		\input{tab/chain_ISING_correct_vote_129_017}
		  \begin{axis} [
      axis lines=left,
      xlabel={$a_i$},
      ylabel={$P(q=1)$},
      ymin=0, ymax=1,
      xmin=-2, xmax=2,
      minor x tick num=3,
      minor y tick num=3,
      legend cell align=right,
      legend style={ draw=none, at={(1.3,1.0)}, },
    ]
    \addplot[red]    table[ x index=1, y index=2,  ]                {\plotdata};
    \addplot[gray]   table[ x index=1, y index=3,  forget plot, ]   {\plotdata};
    \addplot[cyan]   table[ x index=1, y index=4,  ]                {\plotdata};
    \addplot[gray]   table[ x index=1, y index=5,  forget plot, ]   {\plotdata};
    \addplot[violet] table[ x index=1, y index=6,  ]                {\plotdata};
    \addplot[gray]   table[ x index=1, y index=7,  forget plot, ]   {\plotdata};
    \addplot[teal]   table[ x index=1, y index=8,  ]                {\plotdata};
    \addplot[gray]   table[ x index=1, y index=9,  forget plot, ]   {\plotdata};
    \addplot[blue]   table[ x index=1, y index=10, ]                {\plotdata};
    \addplot[gray]   table[ x index=1, y index=11, forget plot, ]   {\plotdata};
    \addplot[orange] table[ x index=1, y index=12, ]                {\plotdata};
    \addplot[gray]   table[ x index=1, y index=13, forget plot, ]   {\plotdata};
    \addplot[yellow] table[ x index=1, y index=14, ]                {\plotdata};
    \addplot[gray]   table[ x index=1, y index=15, forget plot, ]   {\plotdata};
    \addplot[magenta]table[ x index=1, y index=16, ]                {\plotdata};
    \addplot[gray]   table[ x index=1, y index=17, forget plot, ]   {\plotdata};
    \addplot[green]  table[ x index=1, y index=18, ]                {\plotdata};
    \legend{-1.00,-0.75,-0.50,-0.25,0.00,0.25,0.50,0.75,1.00}
  \end{axis}
	    \end{tikzpicture}
	}
	\caption{ } \label{fig:IsingSQC129}
    \end{subfigure}
    \caption{Plots of $P(q=1)$ vs. $a_i$ using the Ising model for 12 qubit chains on D-Wave 2X (a) uncorrected and (b) corrected with SQC. (17 different values of $b_{ij}$ were plotted.)  }
    \label{fig:Ising129A}
\end{figure}
\begin{figure}[H]
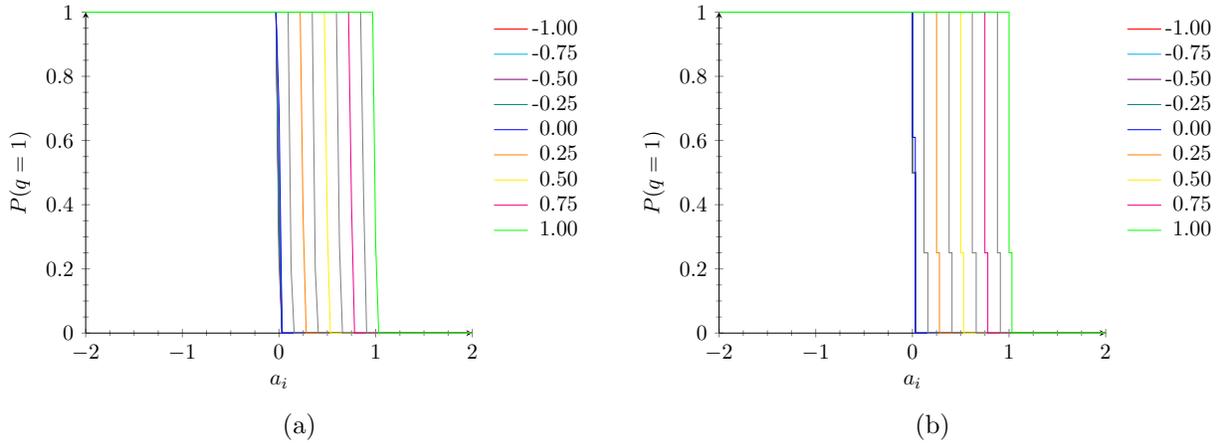

    \begin{subfigure}{0.49\textwidth}
	\scalebox{0.75}{
	    \begin{tikzpicture}
		\input{tab/mqc_ISING_dwave_129_017}
		  \begin{axis} [
      axis lines=left,
      xlabel={$a_i$},
      ylabel={$P(q=1)$},
      ymin=0, ymax=1,
      xmin=-2, xmax=2,
      minor x tick num=3,
      minor y tick num=3,
      legend cell align=right,
      legend style={ draw=none, at={(1.3,1.0)}, },
    ]
    \addplot[red]    table[ x index=1, y index=2,  ]                {\plotdata};
    \addplot[gray]   table[ x index=1, y index=3,  forget plot, ]   {\plotdata};
    \addplot[cyan]   table[ x index=1, y index=4,  ]                {\plotdata};
    \addplot[gray]   table[ x index=1, y index=5,  forget plot, ]   {\plotdata};
    \addplot[violet] table[ x index=1, y index=6,  ]                {\plotdata};
    \addplot[gray]   table[ x index=1, y index=7,  forget plot, ]   {\plotdata};
    \addplot[teal]   table[ x index=1, y index=8,  ]                {\plotdata};
    \addplot[gray]   table[ x index=1, y index=9,  forget plot, ]   {\plotdata};
    \addplot[blue]   table[ x index=1, y index=10, ]                {\plotdata};
    \addplot[gray]   table[ x index=1, y index=11, forget plot, ]   {\plotdata};
    \addplot[orange] table[ x index=1, y index=12, ]                {\plotdata};
    \addplot[gray]   table[ x index=1, y index=13, forget plot, ]   {\plotdata};
    \addplot[yellow] table[ x index=1, y index=14, ]                {\plotdata};
    \addplot[gray]   table[ x index=1, y index=15, forget plot, ]   {\plotdata};
    \addplot[magenta]table[ x index=1, y index=16, ]                {\plotdata};
    \addplot[gray]   table[ x index=1, y index=17, forget plot, ]   {\plotdata};
    \addplot[green]  table[ x index=1, y index=18, ]                {\plotdata};
    \legend{-1.00,-0.75,-0.50,-0.25,0.00,0.25,0.50,0.75,1.00}
  \end{axis}
	    \end{tikzpicture}
	}
	\caption{ } \label{fig:IsingMQC129}
    \end{subfigure}
    \hspace*{\fill} 
    \begin{subfigure}{0.49\textwidth}
	\scalebox{0.75}{
	    \begin{tikzpicture}
		\input{tab/chain_ISING_theoretic_vote_129_017}
		  \begin{axis} [
      axis lines=left,
      xlabel={$a_i$},
      ylabel={$P(q=1)$},
      ymin=0, ymax=1,
      xmin=-2, xmax=2,
      minor x tick num=3,
      minor y tick num=3,
      legend cell align=right,
      legend style={ draw=none, at={(1.3,1.0)}, },
    ]
    \addplot[red]    table[ x index=1, y index=2,  ]                {\plotdata};
    \addplot[gray]   table[ x index=1, y index=3,  forget plot, ]   {\plotdata};
    \addplot[cyan]   table[ x index=1, y index=4,  ]                {\plotdata};
    \addplot[gray]   table[ x index=1, y index=5,  forget plot, ]   {\plotdata};
    \addplot[violet] table[ x index=1, y index=6,  ]                {\plotdata};
    \addplot[gray]   table[ x index=1, y index=7,  forget plot, ]   {\plotdata};
    \addplot[teal]   table[ x index=1, y index=8,  ]                {\plotdata};
    \addplot[gray]   table[ x index=1, y index=9,  forget plot, ]   {\plotdata};
    \addplot[blue]   table[ x index=1, y index=10, ]                {\plotdata};
    \addplot[gray]   table[ x index=1, y index=11, forget plot, ]   {\plotdata};
    \addplot[orange] table[ x index=1, y index=12, ]                {\plotdata};
    \addplot[gray]   table[ x index=1, y index=13, forget plot, ]   {\plotdata};
    \addplot[yellow] table[ x index=1, y index=14, ]                {\plotdata};
    \addplot[gray]   table[ x index=1, y index=15, forget plot, ]   {\plotdata};
    \addplot[magenta]table[ x index=1, y index=16, ]                {\plotdata};
    \addplot[gray]   table[ x index=1, y index=17, forget plot, ]   {\plotdata};
    \addplot[green]  table[ x index=1, y index=18, ]                {\plotdata};
    \legend{-1.00,-0.75,-0.50,-0.25,0.00,0.25,0.50,0.75,1.00}
  \end{axis}
	    \end{tikzpicture}
	}
	\caption{ } \label{fig:IsingTheo129}
    \end{subfigure}
    \caption{Plots of $P(q=1)$ vs. $a_i$ using the Ising model for 12 qubit chains on D-Wave 2X (a) corrected with MQC and (b) theoretically correct. (17 different values of $b_{ij}$ were plotted.)  }
    \label{fig:Ising129B}
\end{figure}

For the sake of this analysis, voting is used to determined if a virtual qubit is true or false.
If the number of physical qubits that are true is greater-than or equal-to the number
of qubits that are false, then the virtual qubit is determined to be true.
Figures \ref{fig:Ising129A} and \ref{fig:Ising129B} are plots of the probability that a virtual qubit
is true verse the value of the qubit and coupler coefficients used in the virtual qubits.
All the physical qubits of a virtual qubit will have the same qubit coefficient, $a_i$, 
and the same coupler coefficient, $b_{ij}$.
Since the coefficients represent a 2 dimensional space, the plots are families of curves.
Each curve represents a single coupler coefficient $b_{ij}$. 
Each curve is a plot of probability vs qubit coefficients $a_i$.

Note that lack of sharp structure in the plots of figure \ref{fig:Ising129A} is indicative of thermal noise.
Though the D-Wave runs at 12 milli-Kelvin there is sufficient thermal noise to be visible in the plots of
qubit behavior (see \citep{Dorband16}).
This is true for the SQC corrected samples as well as the uncorrected samples.
But the plots of figure \ref{fig:Ising129B} are distinctly sharper.
Figure \ref{fig:IsingTheo129} is the plot of a theoretically perfect D-Wave at absolute zero.
Clearly this would be impossible to implement in real hardware.
However MQC does appear to be able to realize this through post processing of D-Wave samples (error correction).
Note that figure \ref{fig:IsingMQC129} and \ref{fig:IsingTheo129} are nearly identical.
The energy of each case of the MQC plotted data were calculated and had the same ground state energy 
as the theoretically perfect D-wave for the same case.
The differences are due to the fact that the theoretical results exhaustively calculated all possible
ground state results giving a more detailed plot. 
MQC only used 40 virtual qubits to calculate the statistics of its plots thus giving not as great detail.

\subsection{Example 3: Chimera Boltzmann machine}\label{sec:ChimeraBoltzM}

A Chimera Boltzmann machine neural network with 2048 neurons was formed by mapping the neurons onto the 2048
qubits of a D-Wave 2000Q in Burnaby, BC. 
It was trained on 100 MNIST images (28x28 pixels) and tested with another 100 MNIST images.
The neural network consisted of three layers, an input layer of size 784,
a hidden layer of 2048 neurons, and an output layer of size 10.
This may not be a realistic artificial neural network, but its intent is to test this style of algorithm 
on the D-Wave with sample correction.

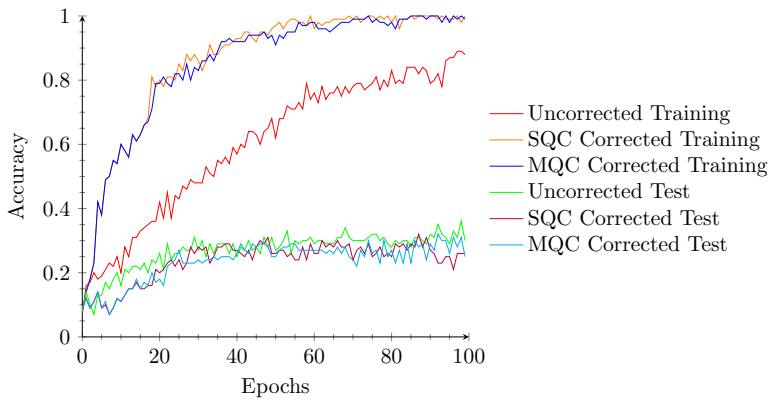
\begin{figure}[H]
\centering
    \begin{subfigure}{0.75\textwidth}
	\scalebox{0.75}{
	    \begin{tikzpicture}
		  \pgfplotstableread{
0 0.07 0.08 0.08 0.08 0.09 0.09
1 0.16 0.14 0.14 0.14 0.12 0.12
2 0.17 0.18 0.18 0.1 0.09 0.09
3 0.2 0.23 0.23 0.07 0.11 0.11
4 0.18 0.42 0.42 0.13 0.14 0.14
5 0.19 0.38 0.38 0.13 0.09 0.09
6 0.21 0.49 0.49 0.17 0.1 0.11
7 0.23 0.5 0.5 0.15 0.07 0.07
8 0.22 0.55 0.55 0.18 0.09 0.09
9 0.25 0.54 0.54 0.2 0.12 0.12
10 0.2 0.6 0.6 0.16 0.11 0.11
11 0.28 0.58 0.58 0.21 0.13 0.13
12 0.25 0.56 0.56 0.2 0.15 0.15
13 0.31 0.63 0.63 0.22 0.15 0.15
14 0.31 0.61 0.61 0.22 0.17 0.18
15 0.33 0.63 0.63 0.21 0.15 0.15
16 0.34 0.66 0.66 0.23 0.15 0.17
17 0.35 0.68 0.67 0.2 0.16 0.16
18 0.36 0.81 0.71 0.24 0.16 0.2
19 0.36 0.78 0.79 0.23 0.21 0.17
20 0.42 0.8 0.79 0.26 0.2 0.18
21 0.37 0.78 0.81 0.21 0.21 0.16
22 0.45 0.81 0.79 0.29 0.23 0.22
23 0.37 0.81 0.78 0.24 0.24 0.23
24 0.44 0.8 0.82 0.26 0.22 0.24
25 0.43 0.85 0.82 0.26 0.24 0.27
26 0.47 0.83 0.8 0.28 0.21 0.23
27 0.46 0.88 0.85 0.27 0.23 0.23
28 0.49 0.86 0.8 0.27 0.28 0.23
29 0.48 0.88 0.84 0.31 0.26 0.23
30 0.48 0.86 0.83 0.27 0.28 0.24
31 0.48 0.83 0.86 0.3 0.27 0.23
32 0.53 0.86 0.86 0.25 0.28 0.24
33 0.51 0.91 0.88 0.29 0.23 0.24
34 0.5 0.88 0.86 0.26 0.25 0.25
35 0.55 0.88 0.89 0.29 0.28 0.24
36 0.54 0.9 0.92 0.29 0.28 0.25
37 0.56 0.91 0.92 0.29 0.29 0.25
38 0.54 0.91 0.93 0.26 0.29 0.25
39 0.59 0.92 0.92 0.29 0.27 0.24
40 0.57 0.93 0.92 0.25 0.27 0.26
41 0.6 0.93 0.92 0.28 0.26 0.29
42 0.59 0.95 0.92 0.27 0.25 0.27
43 0.64 0.95 0.94 0.3 0.27 0.26
44 0.64 0.93 0.94 0.27 0.24 0.29
45 0.63 0.92 0.94 0.28 0.28 0.28
46 0.6 0.95 0.94 0.26 0.3 0.29
47 0.64 0.95 0.95 0.31 0.29 0.29
48 0.65 0.94 0.93 0.28 0.31 0.28
49 0.68 0.96 0.94 0.31 0.26 0.25
50 0.62 0.97 0.91 0.27 0.26 0.25
51 0.68 0.98 0.94 0.28 0.27 0.27
52 0.68 0.96 0.93 0.3 0.27 0.28
53 0.72 0.98 0.95 0.33 0.25 0.28
54 0.71 0.99 0.95 0.26 0.26 0.29
55 0.71 0.99 0.95 0.3 0.24 0.29
56 0.73 0.98 0.98 0.29 0.24 0.27
57 0.71 0.98 0.97 0.3 0.28 0.27
58 0.79 0.97 0.97 0.3 0.24 0.27
59 0.74 1 0.98 0.31 0.29 0.27
60 0.76 0.96 0.98 0.29 0.26 0.27
61 0.73 0.98 0.96 0.3 0.26 0.27
62 0.78 0.97 0.96 0.3 0.3 0.27
63 0.74 0.98 0.96 0.29 0.28 0.26
64 0.76 0.98 0.95 0.29 0.29 0.27
65 0.76 0.99 0.96 0.29 0.28 0.26
66 0.78 0.99 0.97 0.31 0.3 0.28
67 0.75 0.99 0.98 0.31 0.28 0.26
68 0.78 0.99 0.98 0.34 0.28 0.28
69 0.76 1 0.98 0.31 0.29 0.25
70 0.78 0.99 0.99 0.3 0.24 0.24
71 0.79 1 0.99 0.3 0.27 0.22
72 0.79 0.98 0.99 0.3 0.28 0.27
73 0.77 0.99 0.99 0.3 0.26 0.25
74 0.78 1 1 0.31 0.29 0.3
75 0.79 1 0.98 0.32 0.25 0.26
76 0.81 0.99 0.99 0.32 0.26 0.27
77 0.77 1 0.98 0.3 0.28 0.23
78 0.8 0.99 0.98 0.31 0.25 0.3
79 0.78 1 0.97 0.29 0.26 0.25
80 0.83 0.98 0.98 0.28 0.25 0.27
81 0.79 1 0.96 0.3 0.29 0.23
82 0.8 0.96 0.99 0.29 0.28 0.27
83 0.79 0.99 0.99 0.3 0.29 0.23
84 0.84 0.99 0.99 0.3 0.27 0.29
85 0.84 1 1 0.28 0.29 0.23
86 0.84 0.99 1 0.31 0.28 0.3
87 0.82 1 1 0.31 0.32 0.28
88 0.84 1 1 0.3 0.28 0.29
89 0.83 1 0.99 0.29 0.31 0.24
90 0.79 1 1 0.32 0.27 0.29
91 0.8 1 1 0.31 0.26 0.27
92 0.82 0.99 1 0.35 0.23 0.32
93 0.78 1 0.98 0.32 0.23 0.3
94 0.86 0.99 1 0.31 0.25 0.3
95 0.87 0.99 0.98 0.29 0.25 0.26
96 0.87 0.99 1 0.33 0.21 0.31
97 0.89 1 0.99 0.31 0.26 0.28
98 0.89 0.98 1 0.36 0.26 0.31
99 0.88 1 0.99 0.3 0.26 0.25
  }\plotdata;
		  \begin{axis} [
      axis lines=left,
      xlabel={Epochs},
      ylabel={Accuracy},
      ymin=0, ymax=1,
      xmin=0, xmax=100,
      minor x tick num=3,
      minor y tick num=3,
      legend cell align=left,
      legend style={ draw=none, at={(1.8,0.75)}, },
    ]
    \addplot[red]    table[ x index=0, y index=1, ]                {\plotdata};
    \addplot[orange] table[ x index=0, y index=2, ]                {\plotdata};
    \addplot[blue]   table[ x index=0, y index=3, ]                {\plotdata};
    \addplot[green]  table[ x index=0, y index=4, ]                {\plotdata};
    \addplot[purple] table[ x index=0, y index=5, ]                {\plotdata};
    \addplot[cyan]   table[ x index=0, y index=6, ]                {\plotdata};
    \legend{ {Uncorrected Training}, {SQC Corrected Training}, {MQC Corrected Training}, {Uncorrected Test}, {SQC Corrected Test}, {MQC Corrected Test} }
  \end{axis}
	    \end{tikzpicture}
	}
    \end{subfigure}
     \caption{ Result of Chimera Boltzmann machine while using uncorrected and corrected results from the D-Wave }
    \label{fig:CBM}
\end{figure}

Figure \ref{fig:CBM} is a plot of how well this neural network learns the training data with 
uncorrected samples, SQC corrected samples, and MQC corrected samples used to compute the neural network
weight updates.
The plots show that the neural network learns more rapidly with SQC or MQC corrected samples than 
with uncorrected samples.
The plots also show that all of the methods do equally poorly at recognizing non-trained data
which is an indication that due to the use of so many neurons in the hidden layer the neural
network is over-fitting the training data.
SQC does as well as MQC in training, probably due to the fact that D-Wave samples from a single set
of neural network weights are corrected to a single minimal sample for both SQC and MQC 
(see \citep{Dorband17}).
The reason for this requires further study.

\section{Conclusion}\label{sec:Conclusion}

A method (MQC) has been presented that improves the results returned by the D-Wave the vast majority of the time.
It can only be argued at this time that it improves the result, not necessarily corrects it.
It is simple to implement. 
On a 2.67Ghz Intel i7, 10,000 samples are reduced to an optimal sample in 3.87 seconds
and 1,000 samples in 0.44 seconds.
As compared to the current time it takes to get a result from a remote D-Wave of about 2 seconds for 1000 samples
and 6 to 8 seconds for 10,000 samples, it is relatively fast. 
MQC was shown in section \ref{sec:ChimeraBoltzM} to perform as well as SQC on a Chimera Boltzmann machine 
and much better than the D-Wave alone in terms of training rate.
In \ref{sec:VirtualQubit}, MQC demonstrated in a problem that can be computed classically 
to correct the D-Wave results to the global minimum of the objective function in all cases tried.
And in \ref{sec:RandomCoeff}, for a problem that is not feasible to compute classically, applying MQC to a 
more general use of the objective function (i.e. random coefficients),
it improved the results returned by the D-Wave in nearly all cases (99.6\%).
Though it can not be proven that it always obtains the global minimum of the objective function,
it has shown that it reliably generated a more minimum solution than the D-Wave alone.
MQC has been demonstrated to be a useful form of quantum error correction for 
an adiabatic quantum annealing computer.

\section*{Acknowledgment}

The author would like to thank Michael Little and Marjorie Cole of the NASA Advanced Information Systems 
Technology Office for their continued support for this research effort under grant NNH16ZDA001N-AIST16-0091 
and to the NASA Ames Research Center for providing access to the D-Wave quantum annealing computer. 
In addition, the author thanks the NSF funded Center for Hybrid Multicore Productivity Research and 
D-Wave Systems for their support and access to their computational resources.  

\bibliography{QuantumBib}{}

\begin{thebibliography}{10}
\providecommand{\natexlab}[1]{#1}
\providecommand{\url}[1]{\texttt{#1}}
\expandafter\ifx\csname urlstyle\endcsname\relax
  \providecommand{\doi}[1]{doi: #1}\else
  \providecommand{\doi}{doi: \begingroup \urlstyle{rm}\Url}\fi

\bibitem[{Denchev} et~al.(2016){Denchev}, {Boixo}, {Isakov}, {Ding}, {Babbush},
  {Smelyanskiy}, {Martinis}, and {Neven}]{Denchev15}
V.~S. {Denchev}, S.~{Boixo}, S.~V. {Isakov}, N.~{Ding}, R.~{Babbush},
  V.~{Smelyanskiy}, J.~{Martinis}, and H.~{Neven}.
\newblock {What is the Computational Value of Finite-Range Tunneling?}
\newblock \emph{Physical Review X}, 6\penalty0 (3):\penalty0 031015, July 2016.
\newblock \doi{10.1103/PhysRevX.6.031015}.

\bibitem[{Dorband}(2015)]{Dorband15}
J.~E. {Dorband}.
\newblock {A Boltzmann Machine Implementation for the D-Wave}.
\newblock \emph{12th International Conference on Information Technology - New
  Generations (ITNG)}, pages 703--707, 2015.

\bibitem[{Dorband}(2016)]{Dorband16}
J.~E. {Dorband}.
\newblock {Stochastic Characteristics of Qubits and Qubit chains on the D-Wave
  2X}.
\newblock \emph{ArXiv e-prints}, June 2016.

\bibitem[{Dorband}(2017)]{Dorband17}
J.~E. {Dorband}.
\newblock {Improving the Accuracy of an Adiabatic Quantum Computer}.
\newblock \emph{ArXiv e-prints}, May 2017.

\bibitem[{Farhi} et~al.(2000){Farhi}, {Goldstone}, {Gutmann}, and
  {Sipser}]{Farhi00}
E.~{Farhi}, J.~{Goldstone}, S.~{Gutmann}, and M.~{Sipser}.
\newblock {Quantum Computation by Adiabatic Evolution}.
\newblock \emph{eprint arXiv:quant-ph/0001106}, January 2000.

\bibitem[Inc.(2013)]{Dwave13}
D-Wave~Systems Inc.
\newblock {The D-Wave 2X Quantum Computer: Technology Overview}.
\newblock \url{http://www.dwavesys.com/resources/publications}, 2013.
\newblock [Online; accessed 18-May-2016].

\bibitem[{Pudenz} et~al.(2014){Pudenz}, {Albash}, and {Lidar}]{Pudenz2014}
K.~L. {Pudenz}, T.~{Albash}, and D.~A. {Lidar}.
\newblock {Error-corrected quantum annealing with hundreds of qubits}.
\newblock \emph{Nature Communications}, 5:\penalty0 3243, February 2014.
\newblock \doi{10.1038/ncomms4243}.

\bibitem[{Santoro} and {Tosatti}(2008)]{Giuseppe08}
Giuseppe {Santoro} and Erio {Tosatti}.
\newblock Optimization using quantum mechanics: quantum annealing through
  adiabatic evolution.
\newblock \emph{Journal of Physics A: Mathematical and Theoretical},
  41\penalty0 (20):\penalty0 209801, 2008.
\newblock URL \url{http://stacks.iop.org/1751-8121/41/i=20/a=209801}.

\bibitem[{Vinci} and {Lidar}(2017)]{Vinci2017}
W.~{Vinci} and D.~A. {Lidar}.
\newblock {Scalable effective temperature reduction for quantum annealers via
  nested quantum annealing correction}.
\newblock \emph{ArXiv e-prints}, October 2017.

\bibitem[{Young} et~al.(2013){Young}, {Sarovar}, and {Blume-Kohout}]{Young2013}
K.~C. {Young}, M.~{Sarovar}, and R.~{Blume-Kohout}.
\newblock {Error Suppression and Error Correction in Adiabatic Quantum
  Computation: Techniques and Challenges}.
\newblock \emph{Physical Review X}, 3\penalty0 (4):\penalty0 041013, October
  2013.
\newblock \doi{10.1103/PhysRevX.3.041013}.

\end{thebibliography}

\end{document}